\documentclass{aa}

\usepackage{graphicx}
\usepackage{txfonts}

\usepackage{natbib}

\bibpunct{(}{)}{;}{a}{}{,} 

\begin{document}

   \title{{An X-ray characterization of the central region of the SNR G332.5-5.6}}

   \author{A. E. Su\'arez\inst{1,2}, J. A. Combi\inst{1,2}, J. F. Albacete-Colombo\inst{3}, S. Paron\inst{4}, F. Garc\'ia\inst{1,2}, M. Miceli\inst{5}
         }

\authorrunning{Su\'arez et al.}

\titlerunning{An X-ray characterization of the central region of the SNR G332.5-5.6} 


\institute{Instituto Argentino de Radioastronom\'{\i}a (CCT La Plata, CONICET), C.C.5, (1894) Villa Elisa, Buenos Aires, Argentina.\\
\email{aesuarez@iar-conicet.gov.ar}
\and
Facultad de Ciencias Astron\'omicas y Geof\'{\i}sicas, Universidad Nacional de La Plata, Paseo del Bosque, B1900FWA La Plata, Argentina.\\
\email{jcombi@fcaglp.unlp.edu.ar}
\and
Dto de Investigaci\'on en Ciencias Exactas, Naturales e Ingenier\'ia.
Universidad Nacional de Rio Negro. Don Bosco y Leloir (CP 8500) Viedma, Argentina..\\
\email{albacete.facundo@conicet.gov.ar}
\and
Instituto de Astronom\'ia y F\'isica del Espacio (IAFE), CC 67, Suc. 28, 1428 Buenos Aires, Argentina.\\
\and
 INAF-Observatorio Astronomico di Palermo, P.za del Parlamento 1, I-90134 Palermo, Italy }

   \date{Received 08/06/2015 ; accepted 01/09/2015}

 
  \abstract
        {}
        {We present an X-ray analysis of the central region of supernova remnant (SNR) G332.5$-$5.6 through an exhaustive analysis
        of XMM-Netwon observations with complementary infrared observations. We characterize and discuss the origin 
        of the observed X-ray morphology, which presents a peculiar plane edge over the west side of the central region.
        } 
   {The morphology and spectral properties of the X-ray supernova remnant were studied using a single full frame XMM-Newton observation in the 0.3 to 
   10.0 keV energy band. Archival infrared WISE observations at 8, 12 and 24 $\mu$m were also used to
    investigate the properties of the source and its surroundings at different wavelengths.
   }
        {The results show that the extended X-ray emission is predominantly soft (0.3-1.2 keV) and  peaks around 
        0.5 keV, which shows that it is an extremely soft SNR. X-ray emission correlates very well with central regions of bright radio emission. 
        On the west side the radio/X-ray emission displays a plane-like feature with a terminal wall where strong infrared
         emission is detected. Our spatially resolved X-ray spectral analysis confirms that the emission is dominated by weak 
         atomic emission lines of N, O, Ne, and Fe, all of them undetected in previous X-ray studies. These characteristics
         suggest that the X-ray emission is originated in an optically thin thermal plasma, whose radiation is well fitted by a non-equilibrium ionization collisional plasma (VNEI) X-ray emission model. Our study favors a scenario where G332.5-5.6 is expanding in a medium with an abrupt density change (the wall), likely a dense infrared emitting
         region of dust on the western side of the source.
        }
{}

        \maketitle


\section{Introduction}

The energy released by supernovae (SNs) events is of great importance for our understanding of the physics and chemistry of the interstellar medium (ISM). When supernova remnants interact with ambient gas and dust, shocks compress, heat up, ionize, and dissociate the molecular medium. Depending upon the composition of the surrounding ISM, the SNRs can display a vast range of shapes (e.g., Whiteoak \& Green 1996).

  The southern SNR G332.5-5.6, which exhibits an unusual trident radio appearance (Reynoso et al. 2007; Stupar et al. 2007), is particularly interesting. This SNR lies at intermediate galactic latitude $(l,b)$ = (332.5$^\circ$,$-$5.6$^\circ$); its radio emission is composed of two, almost parallel, outer ridges, and a peculiar central region, which is oriented in the NW-SE direction and presents strong correlation with X-ray emission observed by ROSAT in the soft energy range, between 0.1 and 1.2 keV (Reynoso et al. 2007). Because of its spectral characteristics, SNR G332.5-5.6 has been classified as an unusual 
subclass of thermal composite SNRs. 

Reynoso et al. (2007) studied this remnant using radio observations at different frequencies (1384MHz, 1704MHz, 2496MHz, and  2368MHz) obtained with the Australian Telescope Compact Array (ATCA) and HI $\lambda 21$ cm line observations from the Southern Galactic Plane Survey (McClure-Griffiths et al. 2009). They analyzed the radio morphology, polarization, and the distribution of the radio spectral index of the emission with
these data. As a result, Reynoso et al. found that the source presents an average fractional polarization of $\sim$ 35 percent, and a spectral index $\alpha$ = $-$0.7 $\pm$ 0.2. In addition,  they were able to estimate a distance of 3.4 kpc, which implies a linear size of $\sim $30 pc and a height over the plane of 330 pc, using HI $\lambda 21$ cm line in absorption against
the radio continuum emission. Throughout this paper, a mean distance of 3.4 kpc is assumed.

At the same time, Stupar et al. (2007), carried out radio and optical studies of the SNR and background sources, using multiple optical observations from the South African Astronomical Observatory (SAAO), observations from AAO/UKST $H\alpha$ survey (Parker et al. 2005), and radio data at 843 MHz (Bock et al. 1999), 1384 MHz, 2368 MHz, and 4850 MHz (Griffith et al. 1993). Their results confirmed the nonthermal nature of the source and the presence of shock-heated gas through strong SII lines relative to H$\alpha$, and provided strong correlation between H$\alpha$ and radio emission. A very recent paper by Zhu et al. (2015) presents a study of the X-ray properties of a fraction of the central region of SNR G332.5-5.6 using {\it Suzaku} data, which is not affected by straylight effect in that instrument.

With the unprecedented capabilities of the XMM-Newton telescope, it is now possible to perform high-quality imaging and spatially resolved spectroscopy, which are particularly well suited to the study of SNRs, even more well suited for unusual X-ray morphology SNRs, such as G332.5$-$5.6. In this paper, we present the results of the analysis of X-ray data taken with XMM-Newton, as well as a multiwavelength analysis made with the use of observations available from Two Micron All Sky Survey (2MASS; Skrutskie et al. 2006) and the Wide-field Infrared Survey Explorer (WISE; Wright et al. 2010).

The structure of the paper is as follows: in Sect. 2, we describe the XMM-Newton observations and the data reduction process. In Sect. 3, we present the results of our X-ray data analysis, including X-ray images, spectra, mean photon energy map, and the study of the IR data obtained with WISE. In Sect. 4, we discuss a possible scenario to explain the origin of the observed central morphology, and finally, we summarize our main conclusions in Sect. 5. There is additional information on Fermi data analysis in the appendix.


\section{X-ray observations and data reduction}

The SNR G332.5-5.6 has been observed by XMM-Newton satellite with the European Photon Imaging Camera (EPIC). This detector has two cameras, namely MOS1, MOS2 (Turner et al. 2001) and one PN camera (Struder et al. 2001), all of which operate in full frame mode in the 0.2-15 keV energy range. The observation, Obs-Id 0603220201, was centered at ($\alpha_{\rm J2000.0}$=16h42m55s , $\delta_{\rm J2000.0}$=-54$^\circ$ 31' 00"). It was taken with medium filter, Prime Full Window observation mode, and offset on-axis.

The data were analyzed with the XMM Newton Science Analysis System (SAS) version 13.0.0 and the latest calibrations files. The raw data were processed with {\sc emproc} and {\sc epproc} tasks  to obtain a filtered list of events. We searched for high background periods, which affect 4.9, 3.6, and 6 ks of the observation, for the  MOS1, MOS2, and PN cameras, respectively. Thus, the net exposure time of the observation was slightly reduced to 33.8, 35.2, and 27.1 ksec. To create images, spectra, and light curves, we selected events with {\sc Flag} $=$ 0, and {\sc Patterns} $\leq$ 12 and 4 for MOS and PN cameras, respectively. 


\section{Results}


\subsection {X-ray images}

The SNR emits in the 0.3 to 1.2 keV soft X-ray band. X-ray images were obtained in three different energy bands: soft (0.3-0.5 keV), medium (0.5-0.8 keV), and hard (0.8-1.2 keV). We limited the analysis of XMM-Newton  to MOS2 and pn cameras only. Figure 1 shows an X-ray image of G332.5$-$5.6 taken from MOS2 camera in the soft energy range (0.3-1.0 keV). Data from MOS2 camera were exposure-vignetting corrected, and smoothed with a Gaussian kernel for a bin size of 3 pixels. 

The image reveals that the X-ray morphology of G332.5$-$5.6 is consistent with central extended emission as seen by Reynoso et al. (2007) and Stupar et al. (2007) in both radio and X-ray bands. The central structure is elongated in the NW-SE direction. We observe  a plane-like feature, which resembles a wall seen edge-on, toward the west. While diffuse emission is seen extending toward the east. 

In the medium and hard X-ray energy range, where no X-ray emission of the SNR is detected, the observations present straylight contamination due to an off-axis source. As the SNR's emission is only appreciable at E $\leq$ 1.2 keV, the spatial and spectral analysis of the X-ray data is not affected by this phenomenon. 


\subsection {Mean photon energy map}

    \begin{figure}
      \centering
 \includegraphics[width=8.5cm]{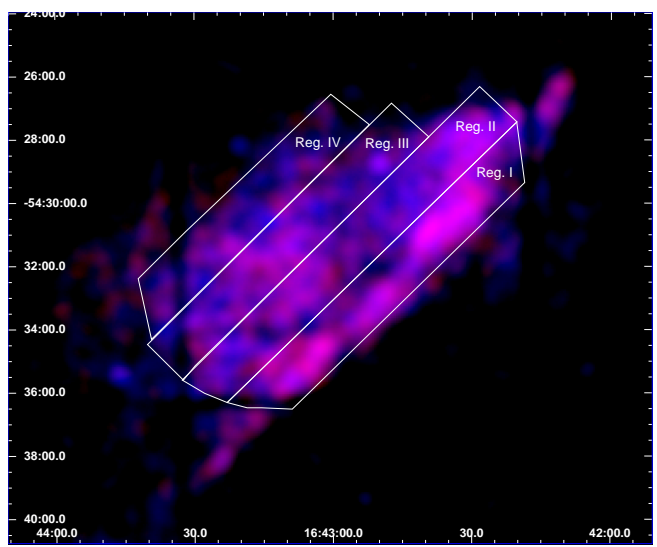}
       \caption{MOS2 image of the inner section of SNR G332.5-5.6 in soft X-ray energy range. Regions with energies between 
       0.3$-$0.6 keV and 0.6$-$1.0 keV are shown in red and blue, respectively. The regions selected 
       for the spectral analysis are also indicated in white. In this image, north is up and east is to the left.}
          \label{Fig1}
    \end{figure}

To study the physical properties of the X-ray emitting gas, the anisotropy of 
the total emission, and to get an insight of the thermal structure toward the central part of the SNR, we computed the mean photon energy (MPE) map (Miceli et al. 2005). 
This MPE map is an image where each computed 
pixel corresponds to the mean energy of the photons detected by PN CCDs in the 0.5$-$1.0~keV 
energy band. Beyond this limit no X-ray emission was detected from the SNR.
It provides information about the spatial distribution of the thermal 
properties of the plasma. Since this characteristic is independent of the spectral model, it cannot be 
corrected by absorption, and the results are biased toward hard X-rays. 

To compute this map, we used the PN event file and created an image with a bin size of $9\arcsec$, which 
 collects a minimum of 8 counts per pixel everywhere in the remnant. For each pixel, we calculated 
the mean energy of the photons by means of an IDL script (Bocchino \& Miceli 2004; Miceli et al 2008) and then we smoothed 
the map by using a Gaussian kernel value of $3\sigma$. The smoothed MPE map of the central region of G332.5-5-6 is shown in Figure 2. According to this map, we are able to confirm that the bulk of the X-ray emission from the SNR has
a mean photon energy of $\sim$0.73$\pm$0.05 keV, whereas a small part at the center of the 
shocked structure has a median energy of  0.65 $\pm$0.04 keV. 
In a few words, the whole X-ray emission is pretty uniform on the extended emission, with an X-ray 
range of emission lower than 0.8 keV. This indicates that the 
post-shocked region at the edge of the SNR had enough time to be uniformly thermalized throughout 
the interaction with the ISM (see next Section). 


\subsection{X-ray spectral analysis}

As a result of the high spatial resolution and sensitivity of the X-ray data, we were able to perform an analysis of the observed extended emission. 
Based on results of our MPE study, first we extracted a single X-ray spectrum of the whole SNR.
For this purpose we made use of the XMM-Newton Extended Source Analysis Software (ESAS) tasks (Snowden et al 2004).
We only took spectra from MOS2 and PN cameras  into account (see Section 3.1). The spectra 
for background emission were subtracted taking into consideration 
circular regions in the different CCDs, where no X-ray emission is detected. Ancillary response files 
(ARFs) and redistribution matrix files (RMFs) were calculated.

    \begin{figure}
\centering
 \includegraphics[width=9cm]{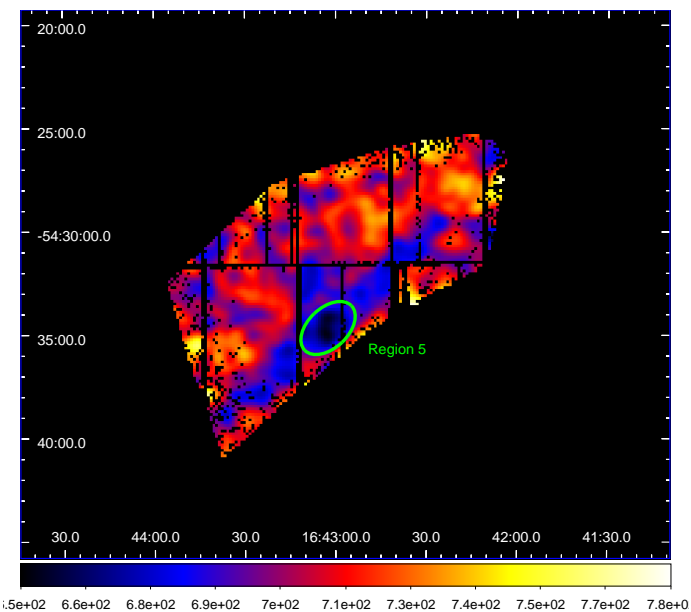}
       \caption{PN mean photon energy map of the 0.5$-$1.0~keV emission 
       (bin size=$9\arcsec$). Pixels with less than 8 counts have been masked out. 
       The color bar has a linear scale and the color-coded energy range is between 
       0.65~keV and 0.78~keV.}
          \label{Fig2}
    \end{figure}

Second,  to analyze the behavior of the X-ray emission, we performed a spatially resolved 
 spectral study of the central emission of the remnant. In this case, we extracted X-ray 
spectra of four regions (as indicated in Fig.\ 1). 
These regions were chosen  to map the change of spectral properties along the SE-NW direction that seems to be the orthogonal direction to the shock front and the dense gas and dust 
structure observed in the IR band (see Section 3.4). In Fig. 3, we show the background subtracted 
X-ray spectra of the four regions indicated in Fig. 1. The spectra are grouped with a minimum 
of 16 counts per bin. Error bars were quoted at 90$\%$ and $\chi^{2}$ statistics were used. The spectral 
analysis was performed using the X-Ray Spectral Fitting Package (XSPEC; Arnaud 1996).

\begin{figure}
\centering
\includegraphics[width=6.3cm,angle=-90]{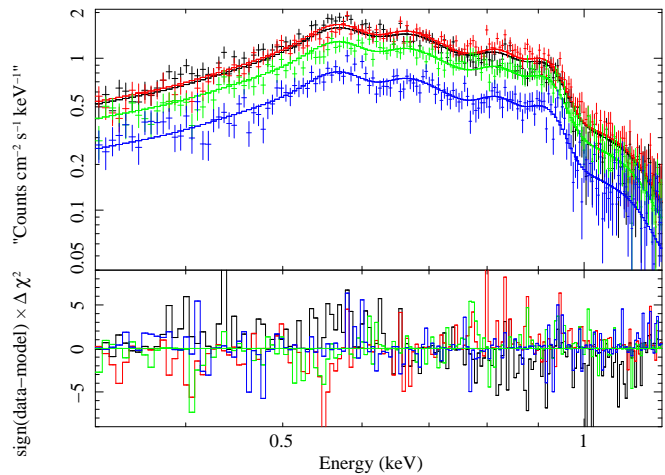}
\caption{X-ray spectra from PN data, color refers to regions 1 (black), 2 (red), 3 (green) and 4 (blue). 
Solid lines indicate the best-fit VNEI model (see Table 1). Lower panels present the $\chi^{2}$ fit residuals.}
\label{Fig3}
\end{figure}

\begin{table*}
\caption{Spectral parameters of the diffuse X-ray emission of the selected regions.}
\renewcommand{\arraystretch}{1.0}
\renewcommand{\tabcolsep}{0.1cm}
\begin{centering}
\begin{tabular}{l|llllllll}
\hline   
\hline  
Model \& Parameters & Total & region 1 & region 2 & region 3 & region 4 & region 5\\ 
\hline
{\bf TBABS*VNEI} & & & & & \\
N$_\mathrm{H}$ [10$^{22}$~cm$^{-2}$]    & 0.26 $\pm$ 0.02       & 0.28 $\pm$ 0.02    & 0.29  $\pm$ 0.03      & 0.25  $\pm$ 0.03      &0.25  $\pm$0.02 &  0.28$\dagger$\\
kT [keV]                                                        & 0.45 $\pm$ 0.07    & 0.42 $\pm$ 0.07       & 0.41  $\pm$ 0.05      & 0.45  $\pm$ 0.07         & 0.54  $\pm$0.14 & 0.25 $\pm$ 0.01\\
N [N$_\odot$]                                                   & 0.57 $\pm$0.24        & 0.59 $\pm$0.21          & 0.40 $\pm$ 0.18       & 0.43  $\pm$ 0.21      & 0.24 $\pm$0.14& 0.59 $\dagger$ \\
O [O$_\odot$]                                           & 0.35 $\pm$ 0.06       & 0.34 $\pm$ 0.04         & 0.28   $\pm$0.04      &  0.32 $\pm$ 0.06      &0.24  $\pm$0.03& 0.34 $\dagger$ \\
Ne [Ne$_\odot$]                                         & 0.48 $\pm$ 0.11         & 0.40 $\pm$ 0.06    & 0.35 $\pm$ 0.05          & 0.46  $\pm$0.08         &0.35 $\pm$0.05& 0.40 $\dagger$\\
Fe [Fe$_\odot$]                                         & 0.40 $\pm$ 0.09       & 0.37 $\pm$ 0.06    & 0.31 $\pm$ 0.04    & 0.36  $\pm$ 0.06      &0.31 $\pm$0.06 & 0.37 $\dagger$\\
$\tau_u$[$10^{10}$~s~cm$^{-3}$]                 &  2.03 $\pm$ 0.48      & 1.70 $\pm$ 0.73         & 2.89   $\pm$ 0.77     & 2.53  $\pm$0.72       & 1.22  $\pm$0.50 & 9.04 $\pm$ 3.04\\
Norm [$10^{-4}$]                                        & 55.84 $\pm$ 18.1      & 24.7 $\pm$ 10.3         & 28.8 $\pm$ 10.6       & 12.1  $\pm$5.11       & 9.36$\pm$3.84 & 6.03 $\pm$ 1.07\\
\hline
$\chi^{2}_{\nu}$ / d.o.f.                               & 1.29/689              & 1.10/188                 & 1.00/189                     & 1.01/191              & 0.99/187& 1.15/226\\
\hline  
Flux(0.3$-$1.0~keV)[$10^{-12}$]                 &30.46                  &  11.35                  & 9.46                  & 5.89                  & 5.31 & 0.13\\
\hline
\end{tabular}
\label{table1}
\tablefoot{Normalization is defined as 10$^{-14}$/4$\pi$D$^2\times \int n_H\,n_e dV$, where $D$ is distance in 
[cm], n$_\mathrm{H}$ is the hydrogen density [cm$^{-3}$], $n_e$ is the electron density [cm$^{-3}$], and $V$ is 
the volume [cm$^{3}$]. Fluxes are absorption-corrected and error values are in the 90\% confidence interval for 
every single parameter, and they are given in units of 10$^{-13}$~erg~cm$^{-2}$~s$^{-1}$. Abundances are 
given relative to the solar values of Anders \& Grevese (1989). $\dagger$ These parameters were fixed to the fit paramenters from Region 1.}
\end{centering}
\end{table*}

The spectra of the regions were fitted by a VNEI emission model, which is compatible with a thin thermal plasma origin. 
The model was multiplicatively affected by the Tuebingen-Boulder ISM absorption model (TBABS; Wilms et al. 2000). We left free 
abundances of N, Ne, O, and Fe. All other element abundances were fixed at their solar values. The X-ray parameters 
of the best fit to the diffuse emission spectra for the different regions are presented in Table 1. 

The spatially resolved spectral analysis shows that the physical conditions of the plasma are clearly homogeneous throughout the central region of the remnant. The mean value of the absorbing column density N$_{\rm H}$ is about 0.26($\pm$0.02) $\times$ 10$^{22}$ cm$^{-2}$. Similarly, the temperature remains at a mean value of 0.45 keV. The uncertainty increase from region 1 to 4 is due to the effect of a lower photon statistic in the X-ray spectra. The values of ionization timescale in all regions are similar, of the order of $\sim$2 $\times 10^{10}$ s cm$^{-3}$. This result is consistent with a nonionization equilibrium state. The relative abundances of the emission are clearly subsolar with a mean of about 0.4 solar value (see Figure 4). These results are consistent with the analysis performed by Zhu et al. (2015) using Suzaku data.
We also found that the absorption corrected flux F$_{\rm X}$ is large in regions where the X-ray brightness is strong, which is due to intrinsic properties of the SNR\ and is not related to the absorbing column material located in the line of sight to the remnant. In this case, we would not expect any link between the N$_{\rm H}$ and the X-ray brightness.  

Additional spectra were taken from the region 5, which is within Region 1, as indicated in green in the MPE map (see Figure 2). These spectra were fitted with a VNEI model, adjusting the abundance parameters to those of Region 1. The resulting fit showed a lower temperature kT=0.25 $\pm$ 0.02, and larger $\tau_u$ = 9.04 $\pm$ 3.0 $\times $10$^{10}$s cm$^{-3}$ than those in Region 1. This is consistent with the shock being slowed down by the interaction with a denser medium. The resulting parameters can be seen in Table 1 in the Region 5 column.

\begin{figure*}
\includegraphics[width=8.6cm,angle=0]{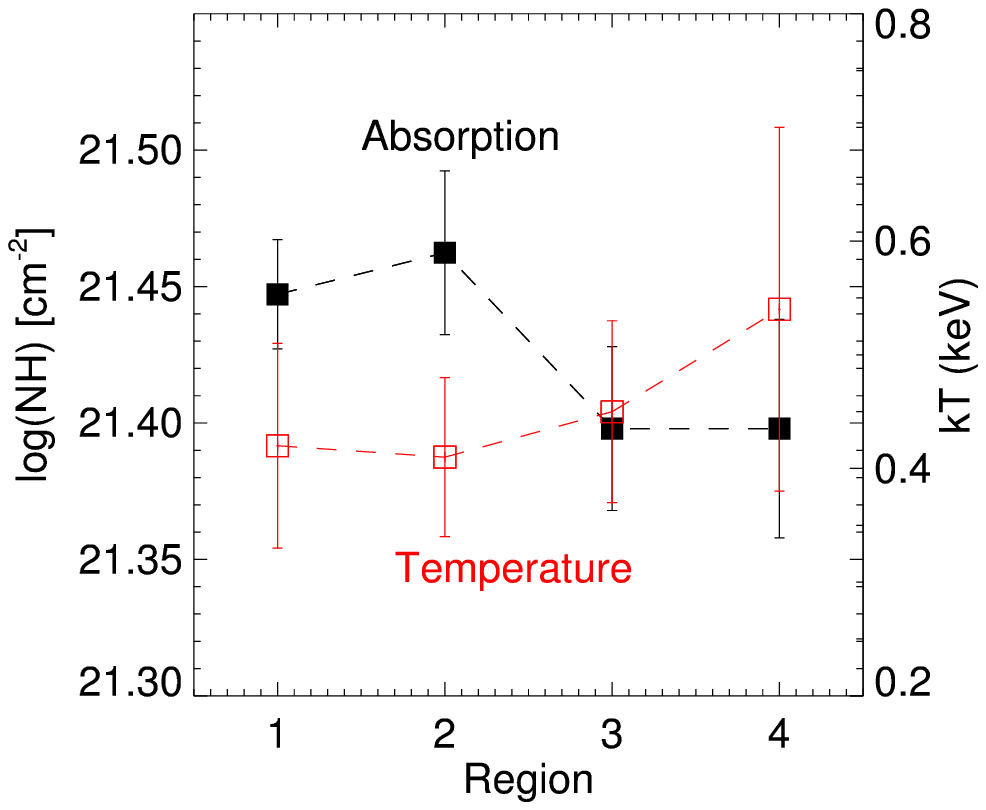}
\includegraphics[width=8.6cm,angle=0]{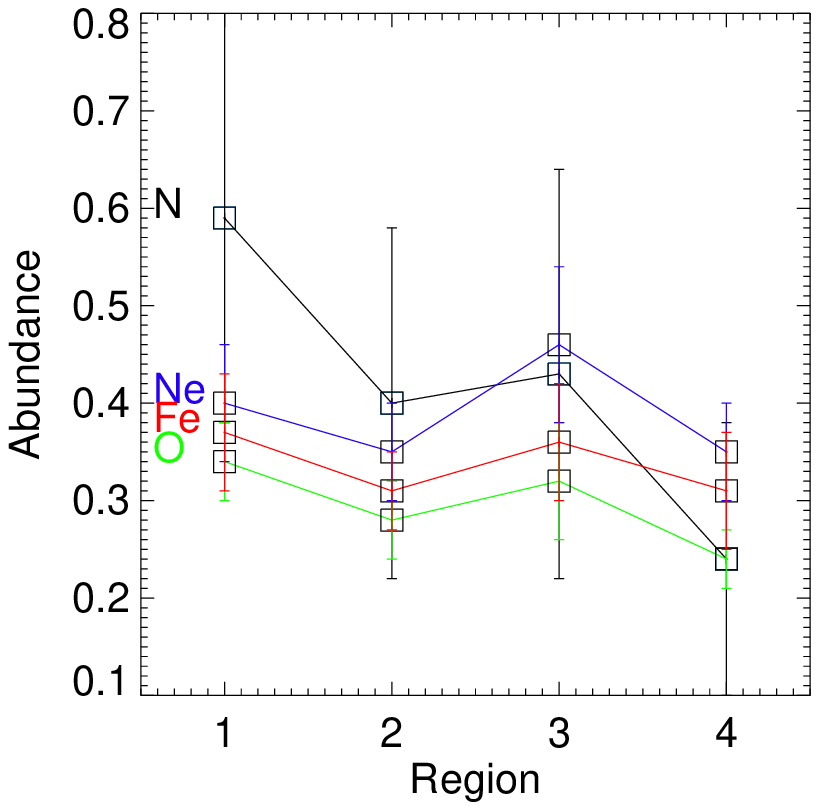}
\caption{X-ray parameters obtained from the spectral analysis as a function of the 
distance from the plane-like region increasing toward the northwest direction. {\it Left:} the absorption-temperature behavior 
seems to be the same inside the error, with a mean of  2.6$\times$10$^{21}$ cm$^{-2}$ and 0.45 keV  for total 
of emission from the SNR. {\it Right:} abundance of N, Ne, Fe, and O are clearly subsolar with a median about 0.4$\pm$0.07.} 
\label{Fig4}
\end{figure*}


\subsection {Analysis of the cold ISM}

Molecular lines and infrared observations toward SNRs can reveal the presence of molecular/dust material, which can affect the SNR expansion. To study the ISM toward the SNR G332.5-5.6, we look for molecular and IR surveys covering this region of the Galaxy. The only data found with adequate angular resolution to perform this kind of study are from WISE at 3.4, 4.6, 12, and 22 $\mu$m. Therefore, we use these data to find any possible connection with the X-ray emission morphology observed by XMM-Newton. The left pane of Fig. 5  shows a composition with WISE bands at 12 and 24 $\mu$m in green and red, and X-ray emission in blue. In contrast, the right panel of Fig. 5 shows IR emission taken by WISE in the bands 4.6, 12, and 24 $\mu$m in blue, green, and red, respectively, with the soft X-ray contours superimposed in yellow. As can be seen from the first image, the X-ray emission (blue) does not display a clear counterpart in the IR bands, which  could be originated in synchrotron emission, shock-heated dust, atomic fine-structure lines, or molecular lines (see, e.g., Reach et al. 2006). However, it is important to note that the plane-like region observed on the western part of the X-ray emission (also observed in the radio bands) coincides with the edge of a bright structure  detected in mid-IR. This sharp edge seen in the X-ray emission can be interpreted as the shock interacting with a denser medium (bright in infrared), which is hampering the expansion of the remnant. This plane-like structure  can be the result of the expansion of the shock in a medium with different possible geometries (i.e., a cylindrical cavity or plain layer), which encounters a denser material. The fact that we  only observe   the interacting part in X-ray may be due to reflected shock produced by the impact with the cloud, which makes the interacting region brightest since it has been shocked and heated twice. The soft X-ray emission, represented in blue in on the left side of Fig. 5, seems to coexist with this IR emission, as seen toward the west. This suggests that the molecular cloud interacting with the SNR is coexisting with the remnant or is in front of it. Assuming that the mid-IR emission arises from dust and polycyclic aromatic hydrocarbon (PAH) molecules, we suggest that the expansion occurred in a less dense medium during the early
evolutionary phase and then the forward shock encountered a more dense medium of gas and dust, as observed at IR wavelengths. To complement this, it would be necessary to observe this region in some molecular lines, such as $^{12}$CO, $^{13}$CO and C$^{18}$O J=1--0, 2--1, and 3--2, to search for the presence of molecular gas, and in case of positive detection, perform a study of its physical conditions.

Finally,  to find if there exists some contribution of a compact central source to the X-ray emission observed on the central part of the SNR, we ran a source detection task {\sc edetect$-$chain} using SAS for the whole X-ray energy range. As a result, we found six point-like X-ray sources, which also present IR emission in WISE data. After a spectral X-ray analysis of these sources we found that none display typical central compact object (CCO) characteristics. In Table 2, the Equatorial coordinates of the sources, which have an IR counterpart, are listed together with the spectral types. The latter are the result of a preliminary photometric analysis based on WISE and 2MASS data. The YSO candidates were determined from the criteria presented by Koenig et al. (2012), based on the magnitudes measured in WISE bands. The  OB-type star candidates were found from a typical color-magnitude diagram using 2MASS bands.

We calculated the threshold counts expected at the PN camera from a putative CCO immersed in the SNR. Considering that the lowest luminosity for a CCO is $\sim$ 10$^{32}$ erg s$^{-1}$ (Halpern et al. 2010), and assuming a distance of 3.4 kpc to the SNR, we obtain a minimum flux for a CCO of $\sim$ 7.6 $\times$10$^{-14}$ erg cm$^2$ s$^{-1}$. Using WebPIMMS tool \footnote{https://heasarc.gsfc.nasa.gov/cgi-bin/Tools/w3pimms/w3pimms.pl}, we can calculate a count rate for this flux for XMM-Newton. Assuming an absorption column density of n$_{H}\sim$ 0.2$\times$10$^{22}$ and a blackbody model source with 0.5 keV temperature, we obtain a count rate for PN camera (with medium filter) of 4.597 $\times$ 10$^{-2}$ cts s$^{-1}$ (counts per second). Taking into consideration that we have a 27.1 ks GTI for PN camera with medium filter, we expect at least $\sim$ 1200 counts for a CCO. In conclusion, as we did not have this kind of detection we can assure with a high level of confidence, that there is no CCO in this region.

    \begin{figure*}
     \centering
 \includegraphics[width=9cm]{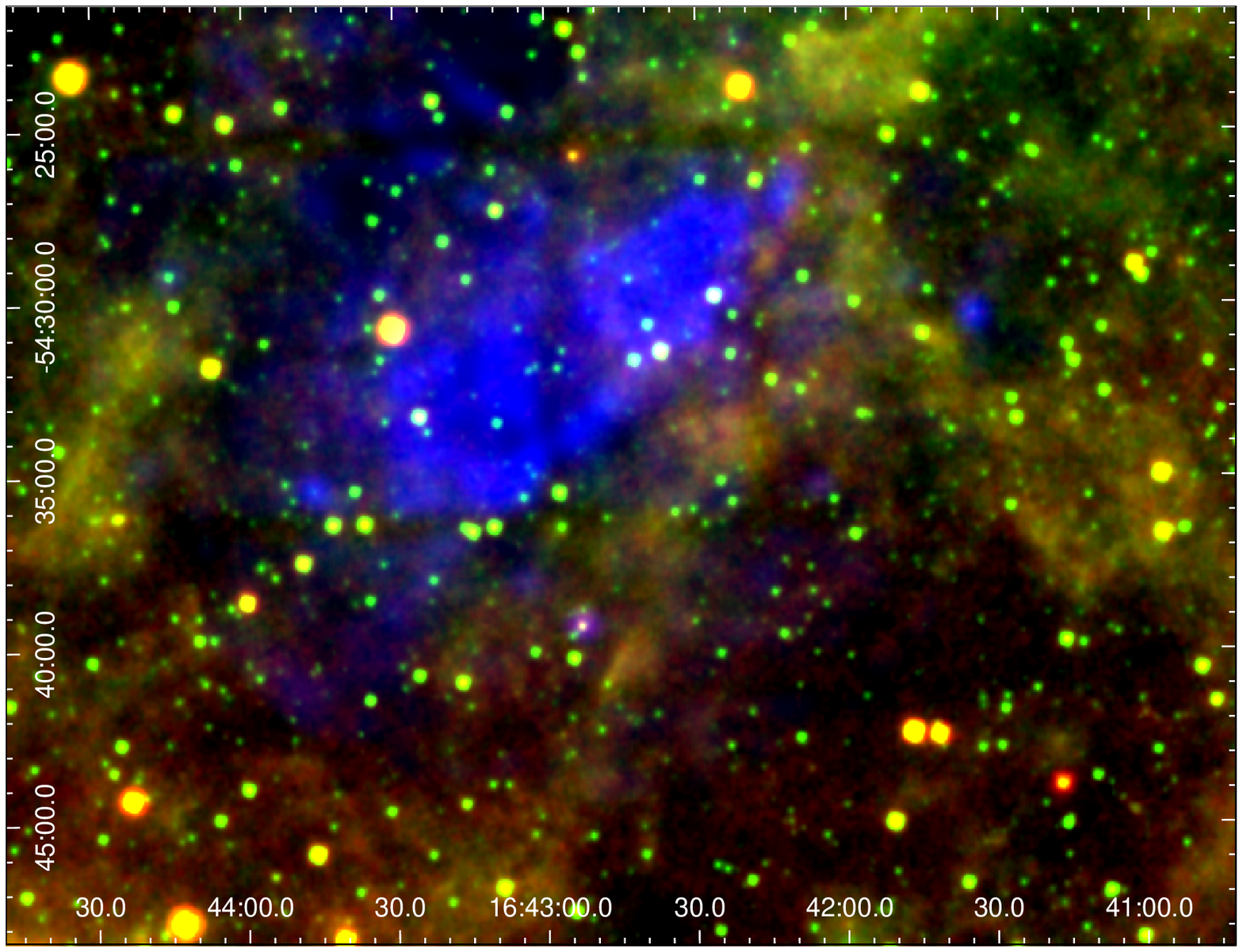}
  \includegraphics[width=8.5cm]{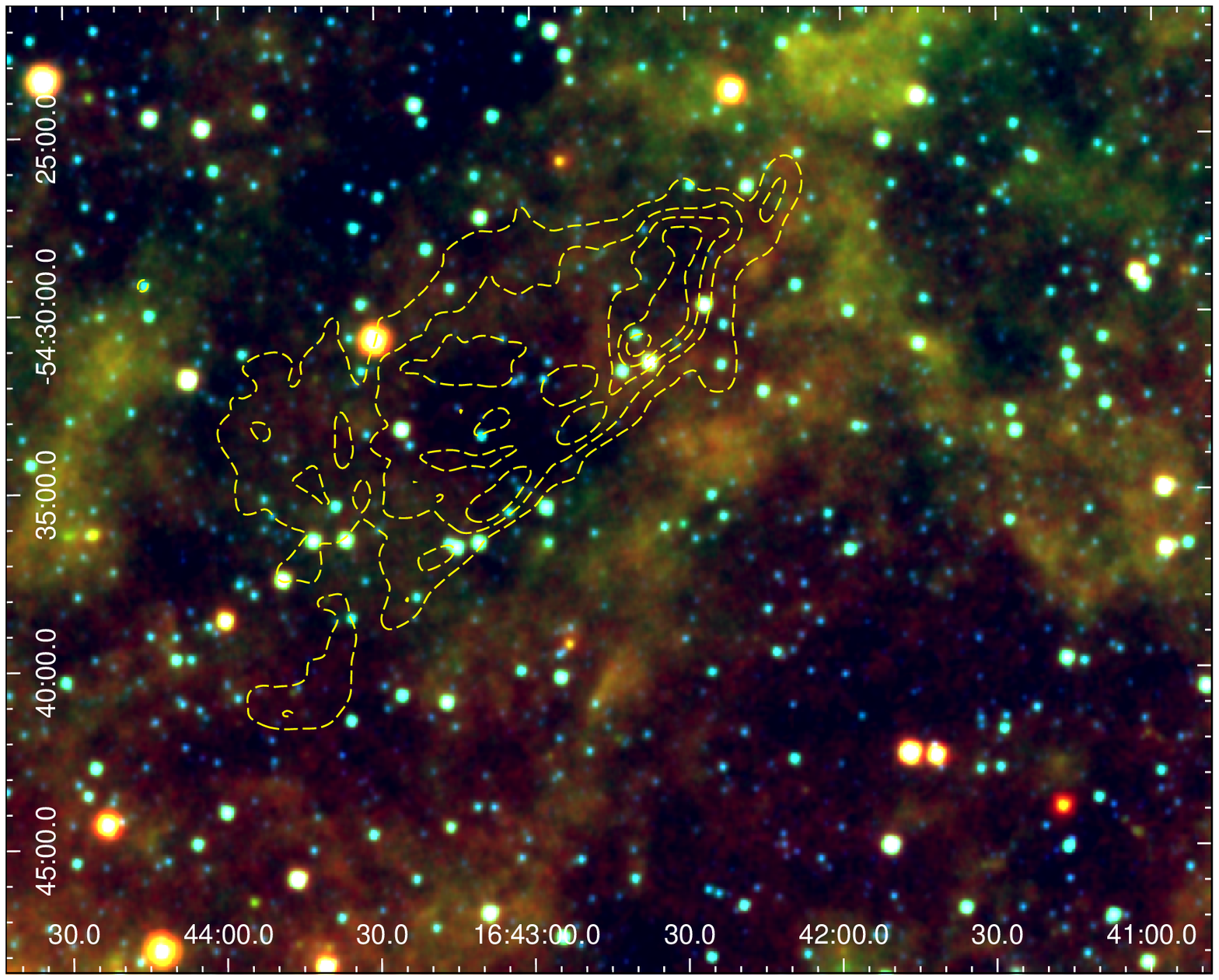}
       \caption{{\it Left panel}: infrared observations of WISE bands at 12 and 24 $\mu$m are shown in green, and red, respectively, with the soft X-ray emission superimposed in blue. {\it Right panel}: infrared observations of WISE bands at 4.6, 12 and 24 $\mu$m are shown in blue, green, and red, respectively.   The soft X-ray contours are superimposed in yellow.}
          \label{Fig5}
    \end{figure*}


\begin{table}
\caption{X-ray sources with IR counterpart}             
\label{table2}      
\centering                          
\begin{tabular}{c c c }        
\hline\hline                 
$\alpha$&$\delta$&Probable Spectral Type\\
\hline
16 41 35.06 &-54 30 17.29,36"&YSO\\
16 42 53.04 &-54 39 19.44,36" & YSO\\
16 42 50.40 &-54 28 45.12,36" &O-type\\
16 43 12.24 &-54 33 33.84,36" &B(0-5)\\
16 43 46.32 &-54 35 25.44,36" &B(0-5)\\
16 44 14.88 &-54 29 09.60,36" &O-type\\
\hline                                   
\end{tabular}
\tablefoot{List of X-ray sources with IR counterpart. The first two columns have the equatorial coordinates of the sources and the third the spectral type.}
\end{table}

We found evidence of dense material associated with IR emission on the west part of the central region of the SNR. In this case, the passage of the shock front could have interacted with interstellar material generating gamma-ray emission. To search for evidence of this possibility, Fermi LAT data were used to search (unsuccessfully) for gamma-ray emission (see App. A for details).


\section {Discussion}

 SNR G332.5-5.6 has been previously classified as a thermal composite or mixed morphology supernova remnant  by Reynoso et al (2007). These kind of remnants are characterized as having radio shells and central X-ray emission, which can peak or be amorphous in reference to the center of the source. This X-ray emission is from thermal origin and product of swept-up interstellar material instead of ejecta. This can be seen by the appearance of lines in the spectra. Parameters like temperature, show uniformity along the remnant, density presents higher values inside the remnant than in the exterior, and metal abundances take  values similar to solar abundances. Another important characteristic is that these kind of remnants are usually seen interacting with molecular or HI clouds (Rho et al 1998). 

In the case of G332.5-5.6 the central X-ray emission has a clear correlation with the central radio emission, however, radio emission is trident-like, not shell-like. Temperature remains almost constant in all regions. A similar behavior can be observed for density parameter, which does not have appreciable changes. Regarding metal abundances, most of metal abundances are solar-like and although N, O, Ne, Fe, present subsolar abundances, they remain near solar values. Infrared data confirm the presence of atomic and molecular/dust material. With this information in mind, we suggest that G332.5-5.6 is an unusual MM subclass category of SNR

The results of the multiwavelength data analysis presented above can be summarized as follows: 

\begin{itemize}

\item The morphology of the extended soft X-ray emission detected by XMM-Newton from the central part of the SNR, is thermal, under nonequilibrium state, and matches very well the emission observed at different radio frequencies. Given the low abundances we can associate the X-ray emitting plasma with shocked ISM and not ejecta.

\item The most striking feature is observed on the western part of the emission, where a plane-like region or interface between two different medium seems to exist.

\item The region to the right part of this interface shows the presence of enhanced IR emission merged with very weak X-ray emission. 

\item No point sources with CCOs characteristics were detected in the central region of the SNR.

\end{itemize}

These results seem to support a scenario where the distorted morphology of the remnant is caused by the expansion of the shock front, first through a low density interstellar medium and then encountering a region with higher density of dust located on the west part of the remnant. 

As the SNR evolves, it compresses and heats up the surrounding medium to X-ray emitting temperatures. When the shock front encounters a high density region, its velocity diminishes and the temperature of the shocked plasma can be considerably lower. In this case, the expansion of the remnant can be significantly impeded, leading to the observed morphology. In some cases, the apparent morphologies of SNRs can depend upon the density and distribution of foreground material. For example, in the case of presence of large values of $N_{\rm H}$ associated with large distances through the ISM, the observed X-ray emission is faint, or the presence of molecular clouds along the line of sight to an SNR  produces excess absorption.

Since the X-ray emission does not originate from the whole SNR, but only from a small part of it (in fact, the radio extension of the SNR is much larger than its X-ray radius), the XMM-Newton observations can only be used to obtain some physical properties of the plasma responsible for the X-ray emission. Therefore assuming that the emitting plasma fills a box with an extension along the line of sight comparable with its extension in the plane of the sky (i.e., $\sim$ 5-6 arcmin), we obtain a volume V of 3.5$\times$ 10$^{57}$ cm$^{3}$ at a mean distance of 3.4 kpc (which is in agreement with the recent studies performed by Zhu et al. 2015) . Using the emission measure (EM) determined from the spectral fitting, we estimated the electron density of the plasma $n_{e}$=$\sqrt{EM/V}$ to be 0.4 cm$^{-3}$. In this case, the density of the nucleons was simply assumed to be the same as that of electrons.

The age, for the X-ray emitting plasma,  $t_{\rm plasma}$ was then determined using the upper limit ionization timescale, $\tau_{ul}$, by $t$= $\tau_{ul}$/$n_{e}$. As a result, the elapsed time after the plasma was heated is $\sim$ 1.4$\times$ 10$^{3}$ yr. The total mass of the X-ray emitting plasma $M_{\rm total}$ can be estimated by

 \begin{equation}
 M_{total} = n_{e} V m_{\rm H} \sim 1.4 M_{\odot}, 
 \end{equation}

 \noindent where $m_{\rm H}$ is the mass of a hydrogen atom.


\section{Conclusions}

In this work, we present a detailed analysis of the X-ray observations obtained with the XMM-Newton telescope of the central region of the SNR G332.5$-$5.6, along with infrared data available in the literature. The results show that the X-ray emission peaks in the soft energy range (0.3-1.0 keV), and confirms that the X-ray morphology has a good correlation with radio emission detected at different frequencies on the central part of the SNR.

From the spatially spectral X-ray analysis, we found that the spectra is thermal and can be well fitted by a VNEI model with a subsolar abundance of N, O, Ne, and Fe. The IR observations suggest that the distorted morphology of the central radio/X-ray emission of the remnant could be the result of the expansion of the shock front, first through a thin interstellar medium and then encounter a region with higher density of dust located on the west of the remnant. All this evidence, suggests that the morphology seen is strongly coupled to the local physical conditions of the ISM.

While the paper by Zhu et al. (2015) provides a global description of the physical properties of the X-ray emitting plasma in (part of) the central region of the remnant, our XMM-Newton data allowed us to perform a spatially resolved spectral analysis, thus revealing the presence of a peculiar region (region 5) characterized by soft X-ray emission, indicative of dense plasma.
{\bf Further} X-ray observations of the central region of this SNR performed by Chandra, at high spatial resolution, are necessary to better understand the nature and morphology of the central part of the SNR.

\begin{acknowledgements}
      
We thank the anonymous referee for her/his insightful comments and constructive suggestions that led to an improved manuscript. A.E.S and F.G. are fellows of CONICET, J.A.C. and J.F.A-C. are researchers of CONICET. J.A.C was supported for different aspects of this work by grants AYA2013-47447-C3-3-P from the Spanish Government, Consejer\'ia de Econom\'ia, Innovaci\'on y Ciencia de Junta de Andaluc\'ia as part of research group FQM-1343, excellence fund FQM-5418, as well as FEDER funds. J.F.A-C was supported by grant PIP 2014-0285 (CONICET). S.P. is supported by CONICET, ANPCyT, and UBA (UBACyT) grants.

\end{acknowledgements}


\begin{appendix}
\section{{\it Fermi} data analysis}

To search any possible gamma-ray emission observed by Fermi telescope, we used $\gamma$-ray data from $Fermi$ Large Area Telescope (LAT) to detect any possible gamma-ray emission in the region centered at the extended X-ray emission ($\alpha$ = 16$^{\rm h} 42^{\rm m} 55\fs0$, $\delta_{\rm J2000.0}$=-54$\degr 31\arcmin 00\farcs0$) within a circle of 4$^{\circ}$ of radius (ROI; region of interest).

The data analysis was performed using the Fermi Science Tools package (v9r32p5) available from the Fermi Science Support Center (FSSC). The data was obtained from the reprocessed Fermi Pass 7 database and the instrumental response function used was the P7REP SOURCE V15 version. We analyzed six years of Fermi data, from August 2008 to August 2014 (2008-08-04T15:43:36 to 2014-08-28T08:34:21, UTC). To prevent the contamination by the Earth's albedo, the events with zenith angle greater than 100$^{\circ}$ or rocking angle greater than 52$^{\circ}$ were filtered. Unbinned likelihood analysis was performed using the $gtlike$ function. To model the background source contributions, we included the entire two-year $Fermi$ Gamma-ray LAT ($2FGL$) catalog of point sources (Nolan et al. 2012) associated with the extended source templates within 4$^{\circ}$ from the ROI center. 

The galactic diffuse background ($gll\_iem\_05.fits$) and the isotropic diffuse background ($iso\_source v05.txt$) were also included in the modeling. All of these background modeling resources are available from the FSSC. Using the full energy range extracted, 100 MeV to 300 GeV, we modeled the differential gamma-ray flux  expected from a point source located at the center of our ROI using a simple power law

\begin{equation}
\frac{{\rm d}N}{{\rm d}E} =N_0 \left(\frac{E}{E_0}\right)^{-\alpha}
,\end{equation}

\noindent where $N_0$ is the normalization factor, $E_0$ is the energy scale, and $\alpha$ the spectral index. 

There is no detection by the $Fermi$-LAT in the chosen energy band at the location of the extended central X-ray source, therefore, only upper limits for the gamma-ray emission  could be determined from this data.  The value found for this upper limit is of 4.7 $\times$ 10$^{-7}$ ph cm$^{-2}$ s$^{-1}$  between 100.0 MeV and 30 GeV.

\end{appendix}

\end{document}